\documentclass[multphys,vecphys]{svmult}

\usepackage{makeidx}         
\usepackage{graphicx}        
\usepackage{multicol}        
\usepackage[bottom]{footmisc}

\makeindex             

\begin{document}

\title*{Leptonic CP Violation and Baryon Asymmetry}
\author{M. N. Rebelo \inst{1}}
\institute{Departamento de F{\'\i}sica and Centro  de F{\'\i}sica
Te{\'o}rica de Part{\'\i}culas (CFTP),\\
Instituto Superior T\'{e}cnico, Av. Rovisco Pais, 1049-001
Lisboa, Portugal
\texttt{rebelo@ist.utl.pt}}
%
%
\maketitle

The observation of neutrino masses leads to the possibility of
leptonic mixing and CP violation. One of the simplest 
extensions of the Standard Model
giving rise to neutrino masses consists of the introduction of one 
righthanded neutrino field per generation, singlet of SU(2).
In the context of the seesaw mechanism this leads to
three light and three heavy neutrinos. The charged current interactions
couple the charged leptons to both the light 
and the heavy physical neutrinos and leptonic CP violation may occur
at low energies as well as at high energies giving rise to the
possibility of leptogenesis.  There are special
scenarios where it is possible
to establish a connection between CP violation 
at the two different scales, an interesting example is included
in this work. Furthermore, we describe how
the conjecture that all phenomena of CP
violation present in nature could have a common origin 
can be realized in the framework of
a further minimal  extension of the Standard Model with CP broken
through the phase of the vacuum expectation value of a complex 
Higgs singlet.

\section{Introduction}
\label{sec:1}
At present there is strong evidence for nonzero neutrino masses
and nontrivial leptonic mixing implying for the first time the 
exitence of physics beyond the Standard Model. In fact in the 
Standard Model (SM) neutrinos are strictly massless and any 
extension giving rise to neutrino masses will contain new ingredients
not present before in the SM. The simplest way of extending
the SM in order to take into account neutrino masses is the 
inclusion of righthanded neutrino singlets, in analogy with all
other fermions in the theory. Yet once righthanded neutrinos are
included both Dirac mass terms and Majorana mass terms for
righthanded neutrinos are allowed. The scale of the Dirac mass terms is
the electroweak scale, v, whilst there are no constraints on the
scale of the righthanded Majorana mass tems. In Grand Unified 
models it is natural to assume this scale, V, to be of the
order of the Grand Unification scale. Mixing and CP violation
in the leptonic sector naturally arise once righthanded neutrinos
are included. In what follows we generally assume their number
to be three although, in fact, the number of righthand neutrino fields
could differ from the number of lefthanded fields. 
When the two scales v and V are very different, with
V much larger than v, the seesaw mechanism \cite{seesaw}
operates providing an elegant explanation for the smallness
of the observed neutrino masses. In the context of seesaw
there are three light neutrinos with small masses and an 
additional number of very heavy neutrinos 
(the number of heavy neutrinos equals the number of righthanded 
neutrinos included) with masses that can be 
of the order of the Grand Unification scale. As a result there can be
leptonic CP violation at low energies as well as at high energies.
Leptonic CP violation at high energies could be the explanation 
for the generation of the observed baryon asymmetry of the
Universe (BAU) via the leptogenesis mechanism 
\cite{Fukugita:1986hr} where a CP asymmetry
generated through the out-of-equilibrium L-violating decays of the
heavy Majorana neutrinos leads to a lepton asymmetry which is
subsequently transformed into a baryon asymmetry by (B+L)-violating
sphaleron processes \cite{Kuzmin:1985mm}. In general there is no
connection between CP violation at low and high energies 
\cite{Rebelo:2002wj} yet this connection can be established 
in special frameworks \cite{varios}. One can go further and 
ask whether there is a framework where all CP violations
have a common origin. In Ref. \cite{Branco:2003rt}
it was shown that this is indeed possible in a small extension of the
Standard Model with neutrino righthanded singlets, a vectorial quark 
isosinglet and a complex Higgs scalar.

\section{Framework}
\label{sec:2}
We work in the context of a minimal extension of the SM 
which consists of adding
to the standard spectrum one right-handed neutrino per generation.
After spontaneous gauge symmetry breaking, the following leptonic mass
terms
can be written:
\begin{eqnarray}
{\cal L}_m  &=& -[\overline{{\nu}_{L}^0} m_D \nu_{R}^0 +
\frac{1}{2} \nu_{R}^{0T} C M_R \nu_{R}^0+
\overline{l_L^0} m_l l_R^0] + h. c. = \nonumber \\
&=& - [\frac{1}{2}  n_{L}^{T} C {\cal M}^* n_L +
\overline{l_L^0} m_l l_R^0 ] + h. c.
\label{lm}
\end{eqnarray}
where $m_D$, $M_R$ and $m_l$ denote the neutrino Dirac mass matrix,
the right-handed neutrino Majorana mass matrix and the charged
lepton mass matrix, respectively, and
$n_L = ({\nu}_{L}^0, {(\nu_R^0)}^c)$ (should be interpreted as a
column matrix).
In order to study CP violation in a weak basis
(WB) it is necessary to consider the
most general CP transformation which leaves the gauge
interaction invariant:
\begin{eqnarray}
{\rm CP} l_L ({\rm CP})^{\dagger}&=&U \gamma^0 {\rm C} \overline{l_L}^T
\quad
{\rm CP} l_R({\rm CP})^{\dagger}=V \gamma^0 {\rm C} \overline{l_R}^T
\nonumber \\
{\rm CP} \nu_L ({\rm CP})^{\dagger}&=&U \gamma^0 {\rm C}
\overline{\nu_L}^T \quad
{\rm CP} \nu_R ({\rm CP})^{\dagger}=W \gamma^0 {\rm C} \overline{\nu_R}^T
\label{cp}
\end{eqnarray}
where U, V, W are unitary matrices acting in flavour space
and where for notation simplicity we have dropped here the
superscript 0 in the fermion fields.
Invariance of the mass terms under the above CP transformation,
requires that the following relations have to be satisfied:
\begin{eqnarray}
W^T M_R W &=&-M_R^*  \label{cpM} \\
U^{\dagger} m_D W&=& {m_D}^*  \label{cpm} \\
U^{\dagger} m_l V&=& {m_l}^* \label{cpml}
\end{eqnarray}
In \cite{bridge},
it was shown, making use of these equations,
that the number of independent CP violating phases which appear in general
in this model is ($n^2 -n$), with n the number of generations. 
The same result was obtained in \cite{onogi}
through an analysis performed in  the physical basis. In the general case
where a Majorana mass term for lefthanded neutrinos is also present
the number of CP violating phases would  be \cite{Branco:gr}
($n^2 + n(n-1)/2$). 

In the case of three generations
(three lefthanded and three righthanded 
 neutrinos), the full neutrino mass matrix, ${\cal M}$
in Eq.~(\ref{lm}), is
$ 6 \times 6$, and has the following form:
\begin{equation}
{\cal M}= \left(\begin{array}{cc}
0 & m \\
m^T & M \end{array}\right) \label{calm}
\end{equation}
We have dropped the subscript in $m_D$ and $M_R$
in order to simply the notation. 
Starting from a weak basis where $m_l$ is already
diagonal and real the neutrino mass matrix is diagonalized by the
transformation:
\begin{equation}
V^T {\cal M}^* V = \cal D \label{dgm}
\end{equation}
where ${\cal D} ={\rm diag.} (m_{\nu_1}, m_{\nu_2}, m_{\nu_3},
M_{\nu_1}, M_{\nu_2}, M_{\nu_3})$,
with $m_{\nu_i}$ and $M_{\nu_i}$ denoting the physical
masses of the light and heavy Majorana neutrinos, respectively. It is
convenient to write $V$ and $\cal D$ in the following form:
\begin{eqnarray}
V&=&\left(\begin{array}{cc}
K & R \\
S & T \end{array}\right) ; \label{matv}\\
{\cal D}&=&\left(\begin{array}{cc}
d & 0 \\
0 & D \end{array}\right) . \label{matd}
\end{eqnarray}
It can be easily verified that both S and R are of order 
$\frac{m}{M}$ (with $R=m T^* D^{-1}$)
and that K is, to an excellent approximation,
the unitary matrix that diagonalizes 
$m_{eff} \equiv  m \frac{1}{M} m^T$:
 \begin{equation}
-K^\dagger m \frac{1}{M} m^T K^* =d \label{14}
\end{equation}
which is the usual seesaw formula. In this approximation 
K is a unitary matrix which coincides with the Maki, 
Nakagawa and Sakata matrix ($V_{MNS}$) \cite{Maki:1962mu}.
The neutrino weak-eigenstates are
related
to the mass eigenstates by:
\begin{equation}
{\nu^0_i}_L= V_{i \alpha} {\nu_{\alpha}}_L=(K, R)
\left(\begin{array}{c}
{\nu_i}_L  \\
{N_i}_L \end{array} \right) \quad \left(\begin{array}{c} i=1,2,3 \\
\alpha=1,2,...6 \end{array} \right)
\label{15}
\end{equation}
and thus the leptonic charged current interactions are given by:\\
\begin{equation}
- \frac{g}{\sqrt{2}} \left( \overline{l_{iL}} \gamma_{\mu} K_{ij}
{\nu_j}_L +
\overline{l_{iL}} \gamma_{\mu} R_{ij} {N_j}_L \right) W^{\mu}+h.c.
\label{16}
\end{equation}
From Eqs.~(\ref{15}), (\ref{16}) it follows that K and R give the
charged current
couplings of charged leptons to the light 
neutrinos $\nu_j$ and to the heavy
neutrinos $N_j$, respectively. In the exact decoupling limit, R
can be neglected and only K is relevant.In this case 
two of the phases that can be factored out of K (in the approximation
of exact unitarity) cannot be rotated
away due to the Majorana character of the neutrino fields and, as a result,
K is left with three CP violating phases (one of Dirac type and two 
of Majorana character). 
However, since we want to
study the connection between CP violation relevant to leptogenesis
and that observable at low energies (e.g., in neutrino oscillations)
we have to keep both K and R.

The present knowledge of leptonic masses and mixing
is still incomplete despite great recent progress.  The 
evidence for solar and atmospheric neutrino oscillations is
now solid and it is already established that the pattern of
the leptonic mixing matrix $V_{MNS}$ is very different from
that of the quark sector ($V_{CKM}$), since only one 
of the leptonic mixing angles, $\theta_{13}$, is small
(the notation is that of the standard parametrization 
of $V_{CKM}$ in \cite{Eidelman:2004wy}). Recent 
KamLAND results \cite{kamland}, a terrestreal long baseline 
experiment which has great sensitivity to 
the square mass difference relevant  for solar oscillations,
$\Delta m^2_{21} $, combined with those of 
SNO \cite{sno} and previous solar
experiments \cite{nuexp} 
lead, for the 1$\sigma$ range \cite{parametros}, to:
\begin{eqnarray}
\Delta m^2_{21} & \equiv & |m^2_{2} - m^2_{1} | =
8.2 ^{+0.3}_{-0.3} \times 10^{-5}\  {\rm eV}^2 \\
\tan ^2 \theta_{12} & = & 0.39 ^{+0.05}_{-0.04} 
\end{eqnarray}
and corresponds to the large mixing angle solution (LMA) of 
the Mikheev, Smirnov and Wolfenstein
(MSW) effect \cite{msw}. On the other hand, atmospheric neutrino 
results from Superkamiokande \cite{atm} and recent important progress by 
K2K \cite{k2k}, which is also a terrestrial long baseline experiment,
are consistent with, for the 1$\sigma$ range \cite{parametros}:
\begin{eqnarray}
\Delta m^2_{32}  & \equiv & |m^2_{3} - m^2_{2} | =
2.2 ^{+0.6}_{-0.4} \times 10^{-3}\  {\rm eV}^2 \\
\tan ^2 \theta_{23} & = & 1.0 ^{+0.35}_{-0.26} 
\end{eqnarray} 
Assuming the range for $ \Delta m^2_{32} $
from SuperKamiokande and K2K, 
the present bounds for  $\sin ^2  \theta_{13} $ from the 
CHOOZ experiment \cite{chooz} at 3$\sigma$ lie \cite{parametros}
in $\sin ^2  \theta_{13} < 0.05-0.07 $.
The  value for the angle
$ \theta_{13} $ is critical for the prospects of detection 
of low energy leptonic CP violation, mediated through a Dirac-type
phase, $\delta$, whose strength is given by ${\cal J}_{CP}$:
\begin{equation}
{\cal J}_{CP} \equiv {\rm Im}\left[\,(V_{11} V_{22}
{V_{12}}^\ast {V_{21}}^\ast\,\right] 
= \frac{1}{8} \sin(2\,\theta_{12}) \sin(2\,\theta_{13}) \sin(2\,\theta_{23})
\cos(\theta_{13})\sin \delta\,, \label{Jgen1}
\end{equation}
Direct kinematic limits on neutrino masses \cite{dir} 
from Mainz and Troitsk and
neutrinoless double beta decay experiments \cite{nonu}
when combined with the given square mass differences 
exclude light neutrino masses higher than order 1~eV. 
Non-vanishing light neutrino masses also have an 
important impact in
cosmology. Recent data from  the Wilkinson Microwave
Anisotropy Probe, WMAP \cite{wmap1}, \cite{wmap2}, 
together with other data, put an upper bound on the sum of light
neutrino masses of 0.7 eV.

\section{General Conditions for Leptogenesis}
\label{sec:3}
The lepton-number asymmetry resulting from the decay 
of heavy Majorana neutrinos, $\varepsilon _{N_{j}}$, 
was computed by 
several authors \cite{sym}.  
The evaluation of $\varepsilon _{N_{j}}$, involves
the computation of the interference between the tree level
diagram and one loop diagrams for the decay of 
the heavy Majorana neutrino $N^j$
into charged leptons $l_i^\pm$ ($i$ = e, $\mu$ , $\tau$) 
leading to:
\begin{eqnarray}
\varepsilon _{N_{j}}
= \frac{g^2}{{M_W}^2} \sum_{k \ne j} \left[
{\rm Im} \left((m^\dagger m)_{jk} (m^\dagger m)_{jk} \right)
\frac{1}{16 \pi} \left(I(x_k)+ \frac{\sqrt{x_k}}{1-x_k} \right)
\right]
\frac{1}{(m^\dagger m)_{jj}}   \nonumber \\
= \frac{g^2}{{M_W}^2} \sum_{k \ne j} \left[ (M_k)^2
{\rm Im} \left((R^\dagger R)_{jk} (R^\dagger R)_{jk} \right)
\frac{1}{16 \pi} \left(I(x_k)+ \frac{\sqrt{x_k}}{1-x_k} \right)
\right]
\frac{1}{(R^\dagger R)_{jj}} \nonumber \\
\label{rmy}
\end{eqnarray}
where $M_k$ denote the heavy neutrino masses,
the variable $x_k$
is defined as  $x_k=\frac{{M_k}^2}{{M_j}^2}$ and 
$ I(x_k)=\sqrt{x_k} \left(1+(1+x_k) \log(\frac{x_k}{1+x_k}) \right)$.
From Eq. (\ref{rmy})
it can be seen that the lepton-number
asymmetry is only sensitive to the CP-violating phases
appearing in $m^\dagger m$ in the WB, where $M_R \equiv M $
is diagonal (notice that this combination is insensitive
to rotations of the left-hand neutrinos).
The simplest leptogenesis scenario corresponds to
heavy hierarchical neutrinos 
where $M_1$ is much smaller than $M_2$ and $M_3$.
In this limit only the asymmetry generated by the lightest
heavy neutrino is relevant, due to the existence of
washout processes, and  $\varepsilon _{N_{1}} $ can
be simplified  into:
\begin{equation}
\varepsilon _{N_{1}}\simeq -\frac{3}{16\,\pi v^{2}}\,\left( I_{12}\,
\frac{M_{1}}{M_{2}}+I_{13}\,\frac{M_{1}}{M_{3}}\right) \,,  \label{lepto3}
\end{equation}
where
\begin{equation}
I_{1i}\equiv \frac{\mathrm{Im}\left[ (m^{\dagger }m)_{1i}^{2}\right] }
{(m^{\dagger }\,m)_{11}}\ .  \label{lepto4}
\end{equation}
Thermal leptogenesis is a rather involved thermodynamical 
non-equilibrium process and depends on additional parameters.
In the hierarchical case the baryon asymmetry only depends on four
parameters \cite{Buchmuller}: the mass $M_{1}$ of the lightest heavy
neutrino, together with the corresponding 
CP asymmetry $\varepsilon _{N_{1}}$ in
their decays, as well as the effective neutrino mass $\widetilde{m_{1}}$  
defined as
\begin{equation}
\widetilde{m_{1}}=(m^{\dagger }m)_{11}/M_{1}  \label{mtil}
\end{equation}
in the weak basis where $M$ is diagonal, real and positive and, finally, the
sum of all light neutrino masses squared, 
${\bar{m}}^{2}=m_{1}^{2}+m_{2}^{2}+m_{3}^{2}$. It has been shown that this
sum controls an important class of washout processes.
Successful leptogenesis would require $\varepsilon _{N_{1}} $ 
of order $10^{-8}$,
if washout processes could be neglected, in order to reproduce
the observed ratio of baryons to photons 
\cite{wmap1}:
\begin{equation}
\frac{n_{B}}{n_{\gamma}}= (6.1 ^{+0.3}_{-0.2}) \times 10^{-10}.
\end{equation}   
Leptogenesis is a non-equilibrium process
that takes place at temperatures $T\sim M_{1}$.
This imposes an upper bound on the effective neutrino mass 
$\widetilde{m_{1}}$ given by the ``equilibrium neutrino mass''
\cite{kfb}:
\begin{equation} 
m_{*}=\frac{16\pi ^{5/2}}{3\sqrt{5}}g_{*}^{1/2}\frac{v^{2}}{M_{Pl}}\simeq
10^{-3}\ \mbox{eV}\; ,  \label{enm}
\end{equation}  
where $M_{Pl}$ is the Planck mass ($M_{Pl}=1.2\times 10^{19}$ GeV),
$v=\langle \phi ^{0}\rangle /\sqrt{2}\simeq 174\,$GeV is the weak scale 
and $g_{*}$ is the effective number of relativistic degrees of
freedom in the plasma and equals 106.75 in the SM case.
Yet, it has been shown \cite{bdbp}, \cite{gnrrs} that successful leptogenesis
is possible for $\widetilde{m_{1}} < m_{*}$ as well as
$\widetilde{m_{1}} > m_{*}$, in the range from 
$\sqrt {\Delta m^2_{12}}$ to  $\sqrt {\Delta m^2_{23}}$. 
The square root of the sum
of all neutrino masses squared ${\bar{m}}$ is constrained, in the
case of normal hierarchy, to be below 0.20 eV \cite{bdbp}, 
which corresponds to an upper bound on light neutrino masses
very close to 0.10 eV. This result is sensitive to radiative
corrections which depend on top and Higgs masses as well as on
the treatment of thermal corrections.  
In \cite{gnrrs} a slightly higher value of 0.15 eV is found.
From Eq.~(\ref{lepto3}) a lower bound 
on the lightest heavy neutrino mass $M_{1}$
is derived. Depending on the cosmological scenario, the range for
minimal  $M_{1}$ varies from order $10^7$ Gev to $10^9$ Gev 
\cite{Buchmuller} \cite{gnrrs}.

\section{Weak Basis Invariants and CP Violation}
\label{sec:4}
In this section we present WB invariants which must vanish if CP
invariance holds. Non-vanishing of any of these WB invariants signals
CP violation. Weak basis invariant conditions are very useful since they
allow us to determine whether or not a Lagrangean violates CP
without the need to go to the physical basis. Clearly they can be
very useful for instance in the study of mass models with 
particular textures or symmetries. The strategy to build these 
conditions was first applied in the context of the Standard 
Model \cite{Bernabeu:1986fc}. The starting point are 
Eqs.~(\ref{cpM}) to (\ref{cpml}). The technique proposed
allows to build several different conditions. Different
conditions may be sensitive to different CP violating phases.
Furthermore some of them are identically zero under particular
circumstances. This requires a careful choice of invariants. 

Since leptogenesis only depends on the product $m^\dagger m$ this
combination must appear in the conditions relevant for leptogenesis.
From Eqs.~(\ref{cpm}), (\ref{cpM}), one obtains :
\begin{eqnarray}
W^{\dagger}h W&=& h^* \nonumber  \\
W^{\dagger}H W&=& H^* \label{wh}
\end{eqnarray}
where $h=m^{\dagger}m$, $H=M^{\dagger}M$.
It can be then readily derived, from Eqs.~(\ref{cpM}),
(\ref{wh}), that CP invariance requires \cite{bridge}:
\begin{equation}
I_1 \equiv {\rm Im Tr}[h H M^* h^* M]=0 \label{i1}
\end{equation}
Analogously several other different conditions can be 
derived \cite{bridge}:
\begin{eqnarray}
I_2 &\equiv& {\rm Im Tr}[h H^2 M^* h^* M] =0
\ \
\label {i2l}\\
I_3 &\equiv& {\rm Im Tr}[h H^2 M^* h^* M H] =0 \label{i3l}
\end{eqnarray}
It has been shown \cite{bridge} that if none of the heavy neutrino masses
vanish and furthermore there is no degeneracy among them these
conditions are independent and do not automatically vanish.
Since there are six independent CP violating phases, one may wonder
whether
one can construct other three independent WB invariants, apart from
$I_i$,
which would describe CP violation in the leptonic sector. This is indeed
possible, a simple choice are the WB invariants ${\bar I}_i (i=1,2,3)$,
obtained
from $I_i$, through the substitution of $h$ by ${\bar h}=m^{\dagger} h_l
m$,
where $h_l=m_l {m_l}^{\dagger}$. For example one has:
\begin{equation}
{\bar I_1}={\rm Im Tr}(m^{\dagger}h_l m H M^* m^T {h_l}^* m^* M)
\label{ibar}
\end{equation}
and similarly for $\bar{I_2}, \bar{I_3}$. As it was the case for $I_i$,
CP invariance requires that $\bar{I_i}=0$.

Since low energy physics is sensitive to $m_{eff}$ it is possible to
show that the strength of CP violation at low energies, observable 
for example through neutrino
oscillations, can be obtained from the following low-energy WB invariant:
\begin{eqnarray}
Tr[h_{eff}, h_l]^3=6i \Delta_{21} \Delta_{32} \Delta_{31}
{\rm Im} \{ (h_{eff})_{12}(h_{eff})_{23}(h_{eff})_{31} \} \label{trc}
\end{eqnarray}
where $h_{eff}=m_{eff}{m_{eff}}^{\dagger},\  h_l=m_l {m_l}^{\dagger}$ and
$\Delta_{21}=({m_{\mu}}^2-{m_e}^2)$ with analogous expressions for
$\Delta_{31}$, $\Delta_{32}$.

\section{Relating CP Violation at low energies with CP Violation
required for Leptogenesis}
\label{sec:5}
It is clear from Eq. (\ref{lm}) that it is possible to choose a 
weak basis where the matrices $m_l$ and $M$ are simultaneously diagonal.
In this case all CP violating phases appear in $m$. There is no loss of
generality in parametrizing the Dirac neutrino mass matrix by
\cite{Hashida:1999wh}:
\begin{eqnarray}
m=U Y_{\triangle} \label{mui}
\end{eqnarray}
with $U$ a unitary matrix and $Y_{\triangle}$ a matrix with triangular
form:
\begin{eqnarray}
Y_{\triangle}= \left(\begin{array}{ccc}
y_{1} & 0 & 0 \\
|y_{21}| \exp(i \phi_{21}) & y_{2} & 0 \\
|y_{31}| \exp(i \phi_{31}) & |y_{32}| \exp(i \phi_{32}) & y_{3}
\end{array}
\right) \label{tria}
\end{eqnarray}
where the $y_{i}$ are real. Since $U$ is unitary, it contains in
general
six phases. However, three of these phases can be rephased away through
the
transformation:
\begin{eqnarray}
m \rightarrow P_{\xi} m  \label{mp}
\end{eqnarray}
where $P_{\xi}={\rm diag}\left(\exp(i\xi_1),\exp(i\xi_2),\exp(i\xi_3)
\right)$. In a WB, this corresponds to a simultaneous phase
transformation
of the left-handed charged lepton fields and the 
left-handed neutrino fields. 
Furthermore, $Y_{\triangle}$ defined by Eq.~(\ref{tria}) can be
written as:
\begin{eqnarray}
Y_{\triangle}= {P_{\beta}^\dagger}\ {\hat Y_{\triangle}}\ P_{\beta}
\label{pyp}
\end{eqnarray}
where $P_\beta =diag(1, \exp(i\beta_1), \exp(i\beta_2))$ and
\begin{eqnarray}
{\hat Y_{\triangle}}= \left(\begin{array}{ccc}
y_{1} & 0 & 0 \\
|y_{21}|  & y_{2} & 0 \\
|y_{31}|  & |y_{32}| \exp(i \sigma) & y_{3}
\end{array}
\right) \label{hat}
\end{eqnarray}
with $\sigma =  \phi_{32} -  \phi_{31} +  \phi_{21}$. It follows from 
Eqs.~(\ref{mui}), (\ref{pyp}) that the matrix  m can then be 
written as \cite{bridge}:
\begin{eqnarray}
m={\hat U_{\rho}} P_{\alpha} {\hat Y_{\triangle}} P_{\beta} 
\label{upy}
\end{eqnarray}
where $P_\alpha =diag(1, \exp(i\alpha_1), \exp(i\alpha_2))$ and
${\hat U}_{\rho}$ contains only one phase $\rho$ as, for
example, in the standard parametrization 
of $V_{CKM}$.
Therefore, in this WB, 
where $m_l$ and $M$ 
are diagonal and real,  the phases 
$\rho$, $\alpha_1$, $\alpha_2$,
$\sigma$, $\beta_{1}$, $\beta_{2}$ are the only physical phases 
and can be used to characterize CP violation in this model.
It follows from here that 
leptogenesis is controlled by the phases
$\sigma$, $\beta_{1}$, $\beta_{2}$. If these three phases vanish
there is no possibility of leptogenesis, still the remaining three
phases can be responsible for low energy CP violation
thus it is possible to have
no CP violation at high energies responsible for leptogenesis
and still have leptonic low energy CP violation  \cite{bridge}.
Conversely one may ask whether it is possible to have 
leptogenesis with no low energy CP violation either of Dirac or
Majorana type \cite{Rebelo:2002wj}. 
The answer to this question can be given by going 
to the weak basis where both $m_l$ and $M$ are real and diagonal.
Then from Eq.~(\ref{14}) one can derive:
\begin{equation}
m= i K {\sqrt d} O^c {\sqrt D },
\label{mmm}
\end{equation}
where ${\sqrt d}$ and ${\sqrt D }$ are diagonal real 
matrices such that   ${\sqrt d} {\sqrt d} = d $, 
${\sqrt D } {\sqrt D } = D $ and $O^c$
is an orthogonal complex matrix, i.e.  
$O^c {O^c}^T =  {1\>\!\!\!{\rm I}} $
but in general $O^c {O^c}^\dagger \neq  {1\>\!\!\!{\rm I}} $. 
It is clear that with this parametrization the
product $m^\dagger m$, relevant for leptogenesis, is
insensitive to K. It is also clear from  Eq.~(\ref{14})
that K is insensitive to the matrix $O$.
Yet, although a connection cannot be established in general,
it can be established in special frameworks. 

Here we present an interesting illustrative example of such
a connection \cite{Branco:2002xf}. Starting from the 
parametrization of Eqs. (\ref{mui}) and (\ref{tria})
it follows that $U$
does not play any r\^ ole for leptogenesis since it cancels out in
the product $m^\dagger m$. This suggests the simplifying choice 
of taking $U= {1\>\!\!\!{\rm I}} $. With this choice several
texture zeros were studied for the matrix  $Y_{\triangle}$.
Two patterns with
one additional zero in $Y_{\triangle}$ where found to be
consistent with low energy physics (either with hierarchical 
heavy neutrinos or two-fold quasi degeneracy):
\begin{equation}
\left(\begin{array}{ccc}
 y_{11}   & 0       &0 \\
 y_{21}\,e^{i\,\phi_{21}}  & y_{22}       & 0 \\
0   & y_{32}\,e^{i\,\phi_{32}} & y_{33}
\end{array}\right) \, ,\qquad 
\left(\begin{array}{ccc}
 y_{11}   & 0       & 0 \\
 0  & y_{22}       & 0 \\
 y_{31}\,e^{i\,\phi_{31}}   & y_{32}\,e^{i\,\phi_{32}} & y_{33}
\end{array}\right)
\label{duas}
\end{equation}
Still it is possible to eliminate one of the two remaining
phases and obtain viable leptogenesis together with
specific predictions for low energy physics consistent with the
known experimental constraints. In Ref. \cite{Branco:2002xf}
special examples were built with strong hierarchies
in the entries of  $Y_{\triangle}$
parametrized in terms of powers of a small parameter.

The question of whether the sign of the baryon asymmetry 
of the Universe can be related to CP violation in neutrino 
oscillation experiments was addressed by considering 
models with only two heavy neutrinos \cite{Frampton:2002qc}. 
In this case the Dirac mass
matrix has dimension $3 \times 2 $. The interesting examples 
correspond to textures of the form given above in Eq. (\ref{duas})
with the third column eliminated and corresponds to the
most economical extension of the SM leading to leptogenesis.
In this case the number of parameters is further reduced
and the remaining non zero parameters are strongly constrained
by low energy physics. This fact leads to a definite 
relative sign between Im $(m^\dagger m)^2_{12}$
and $\sin 2 \delta $.

\section{A common Origin for all CP violations}
\label{sec:6}
CP violation has been observed both in the Kaon 
sector \cite{Christenson:1964fg} and 
in the B-sector
\cite{Aubert:2001sp} \cite{Abe:2001xe}. 
The existence of a matter dominated Universe constitutes indirect 
evidence for CP violation. It has been established 
that within the framework of the SM it is not possible 
to generate the observed size of BAU,
due in part to the smallness of CP violation in the SM. This
provides motivation for considering new sources of CP violation
beyond the KM mechanism.

The question of whether it is possible
to find a framework where all these manifestations of CP violation
have a common origin has been addressed in \cite{Branco:2003rt}
in the context of a small extension of the Standard Model and
also in \cite{Achiman:2004qf} in the framework of a SUSY SO(10)
model. In  \cite{Branco:2003rt} a minimal model is proposed
with spontaneous CP violation, where CP breaking both in the
quark and leptonic sectors arises solely from a phase $\alpha$
in the vacuum expectation value of a complex scalar singlet
S, with $\langle S \rangle = \frac{V}{\sqrt{2}} \exp (i \alpha )$. 
Since S is an $ SU(2) \times U(1) \times SU(3)_c $ singlet, 
$V$ can be much larger
than the electroweak breaking scale. Therefore, in this framework
CP violation is generated at a high energy scale. In order for
the phase $\alpha$ to generate a non-trivial phase at low
energies in the Cabibbo, Kobayashi and Maskawa matrix, one
is led to introduce at least one vector-like quark, whose
lefthanded and righthanded components are singlets under $SU(2)$.
In the leptonic sector, righthanded neutrinos play the r\^ ole
of the vector-like quarks, establishing the connection between
CP breaking at high and low energies, and allowing also for the
possibility of leptogenesis. 

The model considered consists of adding to the SM the following fields:
one singlet charge $- \frac{1}{3}$
vectorial quark $D^0$, three righthanded neutrino fields $\nu_{R}^0$
(one per generation) and a neutral scalar singlet field, $S$. 
A $Z_4$ symmetry is imposed, under which the fields 
$D^0$,  $S$, ${\psi _ l^0}$ (the lefthanded lepton doublets), 
$l_R^0$ and  $\nu_{R}^0$ transform non
trivially, all other fields remain invariant 
under the $Z_4$ symmetry.

The scalar potential will contain terms in $\phi$
and S with no phase dependence, together with terms of 
the form $({\mu }^2 + \lambda_1
 S^\ast S +\lambda_2 {\phi ^ \dagger } \phi )(S^2 + S^{\ast 2})$ $ +
\lambda_3 (S^4 + S^{\ast 4})$ which, in general, lead to 
the spontaneous breaking of T and CP invariance  \cite{Bento:1990wv}
with $\phi$ and $S$ acquiring vacuum expectation values (vevs) 
of the form:
\begin{equation}
\langle {\phi}^0 \rangle = \frac{v}{\sqrt 2}, \ \   \ \ \   
\langle S \rangle = \frac{V \exp (i \alpha )}{\sqrt 2}
\label{vev}
\end{equation}
and the  $Z_4$ symmetry is also broken.

After spontaneous symmetry breaking the leptonic mass terms
are given by Eq.~(\ref{lm}). In the model a bare Majorana mass 
term for the righthanded neutrinos would break the 
$Z_4$ symmetry yet, a term of this form is generated through 
the couplings of $\nu_{R}^0$ to the scalar singlet S, 
after $Z_4$ breaking. It was shown in Ref.\cite{Branco:2003rt} that 
leptogenesis is possible in this framework. Furthermore,
whenever the matrix $m^\dagger m$ is real there is also no
CP violation at low energies. On the other hand the matrix
$mm^\dagger$ is always real in this framework. 

In the hadronic sector the phase 
$\delta _{KM}$, generated through spontaneous CP violation
in general is not suppressed  and the  $Z_4$ symmetry
allows to find a solution \cite{Bento:1991ez} of the strong CP problem  
of the type proposed by Nelson \cite{Nelson} and 
Barr \cite{Barr:qx}.

\section*{Acknowledgments}

The author thanks the organizers of 
DARK 2004 for the warm hospitality at Texas A\& M University
and the stimulating Conference. This
work was partially supported by Funda\c{c}\~{a}o para a Ci\^{e}ncia e a
Tecnologia (FCT, Portugal) through the projects,
POCTI/FNU/44409/2002, CFTP-FCT UNIT 777 which 
are partially funded through POCTI (FEDER).

%
%
%

%
%



\printindex
\end{document}